# Image super-resolution reconstruction based on attention mechanism and feature fusion

**Jiawen Lyn [1], Sen Yan [2]**

[1] linj1@tcd.ie
[2] yanse@tcd.ie

**Abstract:** Aiming at the problems that the convolutional neural networks neglect to capture the inherent attributes of natural images and extract features only in a single scale in the field of image super-resolution reconstruction, a network structure based on attention mechanism and multi-scale feature fusion is proposed. By using the attention mechanism, the network can effectively integrate the non-local information and second-order features of the image, so as to improve the feature expression ability of the network. At the same time, the convolution kernel of different scales is used to extract the multi-scale information of the image, so as to preserve the complete information characteristics at different scales. Experimental results show that the proposed method can achieve better performance over other representative super-resolution reconstruction algorithms in objective quantitative metrics and visual quality.

**Keywords:** feature fusion; convolution neural network; image super resolution

## 1. Introduction

With the rapid development of image processing technology, people's requirements for image quality have gradually increased, and how to obtain high-quality images has become an increasingly pressing issue. Single Image Super Resolution (SISR) reconstruction technology aims to recover high resolution (High) from Low Resolution (LR) images

Resolution (HR) image, the technology is widely used in computer vision, medical imaging, satellite image, video security monitoring and other fields.

Currently, there are three main types of image SR methods: interpolation-based methods, reconstruction-based methods, and learning-based methods [1] [2]. With the wide application of deep learning in the field of computer vision, Dong et al. Proposed an end-to-end convolutional neural network (Super Resolution Convolutional Neural Network, SRCNN) [3] to learn the mapping relationship between LR and HR images. SRCNN is the first work to apply a convolutional network to image SR. This network shows the superiority of Convolutional Neural Network (CNN) in feature extraction and representation. This network structure can be seen as an extension of the traditional sparse coding method. Subsequently, Dong et al. Proposed a Fast Super Resolution Convolutional Neural Network (FSRCNN)

[4], the use of deconvolution layers at the end of the network effectively obtains the high-frequency information in the original image. As the depth of convolutional neural networks continues to deepen, the difficulty of network training is also increasing. Therefore, in response to this problem, He et al. Proposed Residual Neural Network (ResNet) [5], and proposed the idea of residual learning in the network to accelerate the convergence of the network. On this basis, Kim et al. Proposed deep convolutional networks (Accurate Image Super-Resolution Using Very Deep Convolutional Networks, VDSR) [6]. Through the use of residual learning, the network uses 20 convolutional layers, Effectively increase the network depth. In order to reduce the amount of network parameters,

Kim et al. Further proposed Deeply-Recursive Convolutional Network (DRCN) [7], which uses a recursive structure to introduce recursive supervision and jump connections. In addition, in order to solve the real-time SR problem, Shi et al. Proposed an efficient sub-pixel convolutional neural network (Efficient Sub-Pixel Convolutional Neural Network, ESPCN) [8], which utilizes the up-sampling method of sub-pixel convolution to recover in real time. HR image. Lee et al. Proposed Enhanced Deep Residual Networks (EDSR) [9], removed Batch Normalization (BN) operations, and increased the output characteristics of each layer on a large scale. By effectively designing the network structure, the convolutional neural network has achieved significant results in image SR that are superior to traditional methods.

However, although the deep learning-based image SR method has achieved relatively good reconstruction results, there are still the following problems: (1) Although the non-local characteristics of the image are widely used in traditional methods [10] [11], most CNN-based SR methods do not fully utilize the inherent non-local similarity features of images; (2) most CNN-based SR methods are mainly focused on designing deeper or wider networks to learn more Of image features, rarely use higher-order feature statistics. The literature [12] shows that the higher-order features of the image are more representative than the first-order features; (3) Most CNN-based SR methods only extract image features at a single scale, ignoring image details at different scales.

In response to the above problems, we propose a network structure based on attention mechanism and multi-scale feature fusion. First, extract the original features of the image through the shallow feature extraction module; then in the attention mechanism and multi-scale feature fusion module, the non-local characteristics of the image and the second-order statistical features are fused through the attention mechanism, and at the same time, through the multi-scale feature extraction module To obtain image feature information at different scales; finally, the extracted depth features are passed through a reconstruction module to realize image reconstruction.2. Methods and Network Structures

*2. Related Work*

*2.1.Non-local block*

The non-local features of images are widely used in the field of traditional image reconstruction. The basic idea is to search for similar image blocks in the entire image, and use the complementary information provided by these similar image blocks to perform image reconstruction. Buades et al. Proposed a method to improve the image denoising ability based on the non-local average operation of all pixels in the image [10]. The main idea is that for a certain image block, search for similar image blocks in the entire image range, and then give different weights to different image blocks according to the similarity between the image blocks, and the weighted average is finally obtained. Noise image block is a non-local image denoising method. However, in the field of deep learning, image non-local methods have not been widely used. Wang et al. Proposed a non-local Neural Networks (NLNN) [13], which can well capture the dependency relationship between pixels at distant locations, thus integrating non-local operations into use. It is used in non-local convolutional neural networks for video classification. Non-local thought is a classic computer vision method, which has been widely used in neural networks.

*2.2. Attention mechanism*

In recent years, attention mechanisms have been widely used in deep neural networks. The NLNN mentioned above is also the application of attention thought.

Hu et al. Proposed SENet [14] to learn the correlation between channels and adaptively correct the corresponding feature strength between channels through the global loss function of the network. The network structure has achieved significant performance improvements in image classification. Woo et al. Proposed the Convolutional Block Attention Module (Convolutional Block Attention

Module, (CBAM) [15] combines the attention mechanism of space and channel. Compared with SENet which only focuses on the attention mechanism of channel, CBAM can achieve better results. Zhang et al. Introduced the attention mechanism into the super-resolution reconstruction task [16], and used the attention mechanism to treat different channels differently, thereby improving the representation ability of the network. However, these networks only pay attention to the first-order features of the image (such as global average pooling) and ignore the higher-order information features of the image. Dai et al. Proved that second-order statistics can better capture image features [17]. Therefore, they designed a second-order attention network (Second-order Attention Network, SAN) by considering higher-order feature statistics To adaptively utilize channel characteristics.

3. Our Method

As shown in Figure 1, our proposed network structure is divided into three parts, shallow feature extraction (shallow feature extraction, referred to as (SF), deep feature extraction based on attention mechanism and multi-scale (Deep feature extraction, DF for short) and reconstruction (reconstruction, RE). Next we will describe each part.

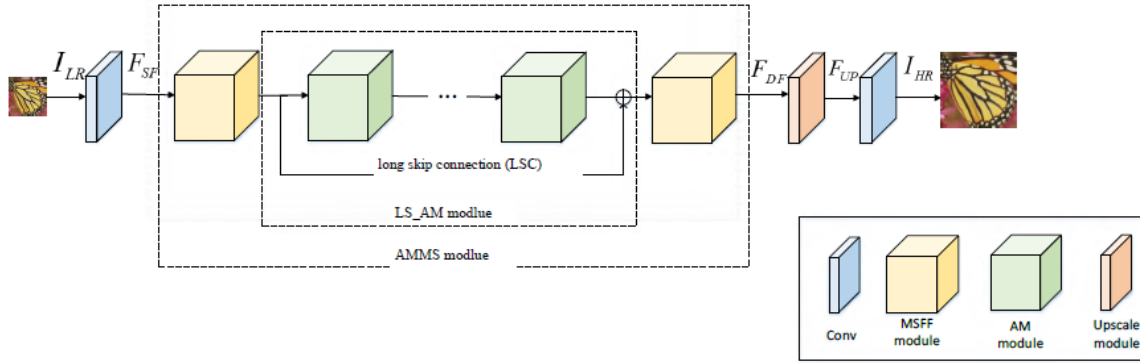

**Figure 1.** The proposed network structure.

. 2.2. Shallow Feature Block

After extracting coarse feature with a convolution layer, we further propose shallow feature block (SFblock) to extract shallow feature and correct the feature in the tail of the network. Our SFblock consists of several 1 x 1 convolution layers and local residual learning.

$$I^{SF} = f_{SF}(I^{LR})$$

. 2.3. D𝑒𝑒p Feature Block

In this part, the extracted shallow feature FSF undergoes deep feature extraction through modules based on attention mechanism and multi-scale feature fusion (AMMS)

$$I^{DF} = f_{AMMS}(I^{SF})$$

where $f_{AMMS}$ means AMMS module operation, for AMMS. The module, as seen in Figure 1, is composed of two Multi-Scale Feature Fusion (MSFF) modules and a LS_AM module. Next we will introduce these two modules separately.

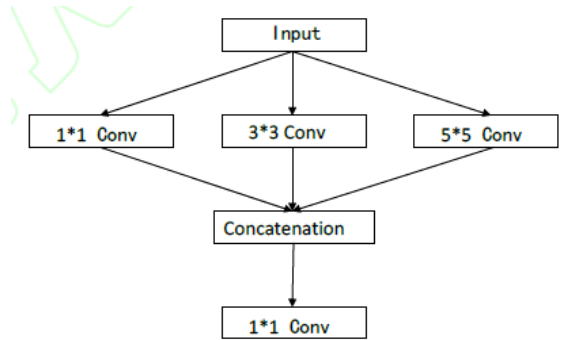

**Figure 2.** Multi-Scale Feature Fusion module.

*2.3.1 Multi-Scale Feature Fusion Modules*

For the multi-scale feature extraction layer, three-scale convolution kernels are used, which are 1×1, 3×3, and 5×5, respectively. Among them, the 1×1 convolution kernel can retain the features of the previous layer and merge with the features of other scales, so that the network contains information features related to shallow features. A nonlinear layer is used after each convolution kernel to improve the nonlinear mapping ability of the module. The specific details are shown in Figure 2.

2.3.2 LS_AM Module

For the LS_AM module, its structure is shown in Figure 3. It consists of M attention mechanism (AM) modules and long skip connection (LSC). Use FAM and FMAM to represent the input and output characteristics of the LS_AM module, respectively.

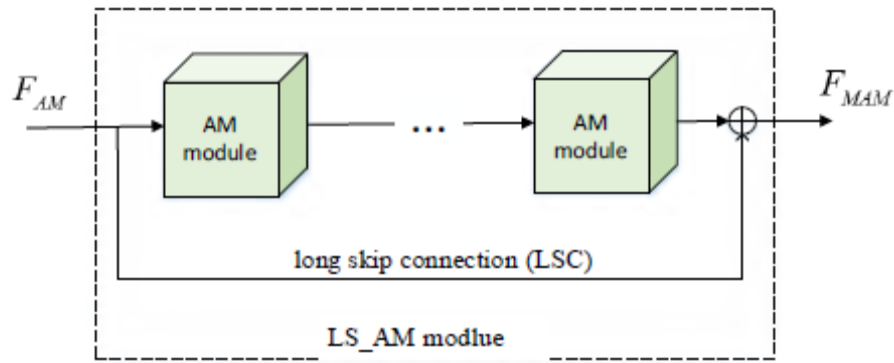

**Figure 3.** LS_AMMS module

2.3.3 LS_AM Module

The attention mechanism (AM) module is shown in Figure 4 and Figure 5. It extracts image features in two ways, including non-local (NL) and second-order (SO) feature extraction, and fuse the extracted features As the output of each attention unit module.

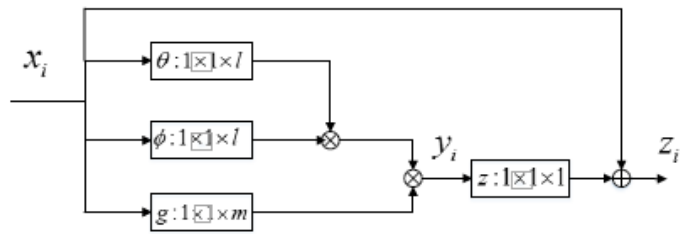

**Figure 4.** Non-local module

Literature [12] shows that higher-order features contain more statistical features than traditional first-order features. The traditional convolution and global average pooling operations can only extract the first-order features of the image. Therefore, inspired by [15], we use global covariance pooling (GCP) to extract the second-order statistical features of the image.

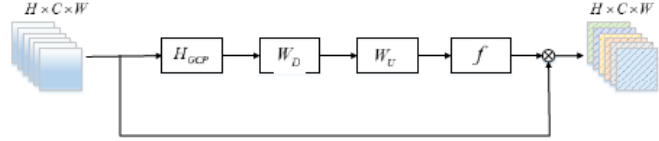

**Figure 5.** Sencon-order module

## 4. Experiments

*4.1. Datasets and Metrics*

We used 91 images from the Timofte dataset [19] and BSD200 [19] from the BSD dataset as the training set. we are at tested on 3 standard data sets, namely: Set5 [20], Set14 [21] and BSD100 [22], these data sets are usually used for image SR reconstruction. The Set5 data set contains images of animals and plants; Set14 contains images of animals, plants and scenes, which have more detailed information than Set5; the Urban100 data set contains images of urban buildings, which have more edge information and are more difficult to reconstruct. In order to compare with other reconstruction methods, we used PSNR and SSIM indicators to evaluate the reconstruction quality. We follow the existing method, only for the luminance channel in the YCbCr color space (Y channel) For super-resolution reconstruction, in order to facilitate the display, the other two chroma channels use the bicubic interpolation algorithm similar to other methods.

*4.2. Training Details*

As for training, we used 16 HR RGB image patches with a size of 192 × 192 randomly cropped from the training images in each training batch. The corresponding LR image for all models with different scale factors (×2, ×3 and ×4) becomes downsampling by adopting the MATLAB function with a bicubic function. During the training process, random vertical flip, random horizontal flip, and 90° rotation augmented patches with a random probability of 0.5. Patch image pixel values are normalize and the average RGB values of the DIV2K [13] dataset which are subtracted from them as pre-processing. We use the Pytorch framework to implement our MLRN and train the model by setting up the ADAM optimizer [19] $\beta_1$= 0.9, $\beta_2$= 0.999, and $\varepsilon$=$10^{-8}$. The training loss function is the L1 loss. The learning rate for all layers is initialized to $1e^{-4}$, halving every 200 epochs, and 1,000 iterations of back-propagation constitute an epoch. Models with different scale factors will be trained from scratch. GPU GTX1080Ti takes about 5 day to train MLRN 1000 epochs

*4.3. Ablation Study*

In order to study the attention mechanism (including non-local and second-order feature extraction operations) and the impact of multi-scale feature fusion modules on the network structure, we introduced the idea of ablation learning. Taking scale factor factor = 4 as an example, we compared the data in Set5 On the set, the impact of the above modules on the average PSNR. It can be seen from Table 1 that the model with the attention mechanism and multi-scale feature fusion has the highest PSNR value.

**Table 1.** Ablation study on effects of non-local, second-order feature extraction operations and multi-scale feature fusion modules. We present the best performance (average PSNR) on Set5 with scale factor ×2 in 1000 epochs.

| Non-local | Second-order | Multi-scale | PSNR |
|---|---|---|---|
| ✓ | × | × | 36.32 |
| × | ✓ | × | 36.78 |
| × | × | ✓ | 36.54 |
| ✓ | ✓ | ✓ | 37.23 |

*4.4. Benchmark Results*

We combine the network structure proposed in this article with Bicubic, SRCNN [3], SCN [24], LapSRN [25] and above 4 image super-resolution methods were compared. The SRCNN model has a three-layer convolution structure, which enlarges a low-resolution image to the target size through bicubic interpolation as the network input. The SCN model combines the traditional sparse coding idea on the basis of the SRCNN three-layer structure. The model has a five-layer convolution structure. The LapSRN model incorporates the idea of the Laplacian pyramid model in the network, which has a 24-layer network structure. Among them, Bicubic method we use Matlab

The implementation of interp2 function, the realization of other comparison methods are from the author's public source code. Table 2 shows the average PSNR of different methods on the Set5, Set14, and BSD100 test sets. Table 3 shows the average value of SSIM for different methods on Set5, Set14, BSD100 test set. We marked the results with the best PSNR and SSIM values in bold. It can be seen from the table that when the amplification factor is 2, our method has higher average PSNR and SSIM on each test data set than several other methods. When the factors are at 3 or 4, we can get the same result. This shows that our network performance has improved significantly compared to these algorithms.

**Table 2.** Public benchmark test results. Average PSNR/SSIMs for scale factor ×2, ×3, and ×4 on datasets Set5, Set14, BSD100, and Urban100.

| Dataset | Scale | Set5 | Set14 | BSD100 | Urban100 |
|---|---|---|---|---|---|
| Bicubic | ×2 | 33.66/0.9299 | 30.24/0.8688 | 29.56/0.8431 | 26.88/0.8403 |
|  | ×3 | 30.39/0.8682 | 27.55/0.7742[1] | 27.21/0.7382 | 24.46/0.7349 |
|  | ×4 | 28.42/0.8104 | 26.00/0.7027 | 25.96/0.6675 | 23.14/0.6577 |
| SRCNN | ×2 | 36.66/0.9542 | 32.42/0.9063 | 31.36/0.8879 | 29.50/0.8946 |
|  | ×3 | 32.75/0.9090 | 29.28/0.8208 | 28.41/0.7863 | 26.24/0.7989 |
|  | ×4 | 30.48/0.8628 | 27.49/0.7503 | 26.90/0.7101 | 24.52/0.7221 |
| VDSR | ×2 | 37.53/0.9587 | 33.03/0.9124 | 31.90/0.8960 | 30.76/0.9140 |
|  | ×3 | 33.66/0.9213 | 29.77/0.8314 | 28.82/0.7976 | 27.14/0.8279 |
|  | ×4 | 31.35/0.8838 | 28.01/0.7674 | 27.29/7251 | 25.18/0.7524 |
| LapSRN | ×2 | 37.52/0.9591 | 33.08/0.9130 | 30.41/0.9101 | 37.27/0.9740 |
|  | ×3 | 33.82/0.9227 | 29.79/0.8320 | 27.07/0.8272 | 32.19/0.9334 |
|  | ×4 | 31.51/0.8855 | 28.19/0.7720 | 25.21/0.7553 | 29.09/0.8893 |
| MemNet | ×2 | 37.78/0.9597 | 33.28/0.9142 | 32.08/0.8978 | 31.31/0.9195 |
|  | ×3 | 34.09/0.9248 | 30.00/0.8350 | 28.96/0.8001 | 27.56/0.8376 |
|  | ×4 | 31.74/0.8893 | 28.26/0.7723 | 27.40/0.7281 | 25.50/0.7630 |
| Our Method | ×2 | **37.92/0.9623** | **33.51/0.9160** | **32.23/0.8997** | **31.88/0.9290** |
|  | ×3 | **34.23/0.9299** | **30.22/0.8369** | **29.01/0.8056** | **27.88/0.8499** |
|  | ×4 | **31.95/0.8912** | **28.43/0.7748** | **27.49/0.7334** | **25.78/0.7753** |

In addition, we selected three pictures from the three test sets of Set5, Set14 and BSD100 for testing. Zoom in on specific areas of the image to better observe the effect of texture detail reconstruction. Figure 5 show the reconstruction effect of this algorithm and several other algorithms at scale factors of 2, 3 and 4, respectively. From the perspective of subjective visual effects, compared with several other methods, the image reconstructed by this method restores more high-frequency details and produces clearer edge effects. Therefore, combining the two factors of subjective effect and objective index, the algorithm in this paper can get better reconstruction effect than the mainstream super-resolution reconstruction algorithm.

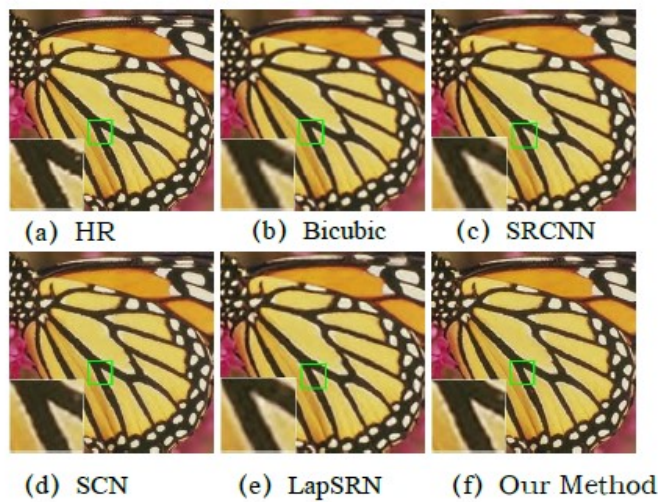

**Figure 5.** Visual quality of different algorithms with the scale factor is 3

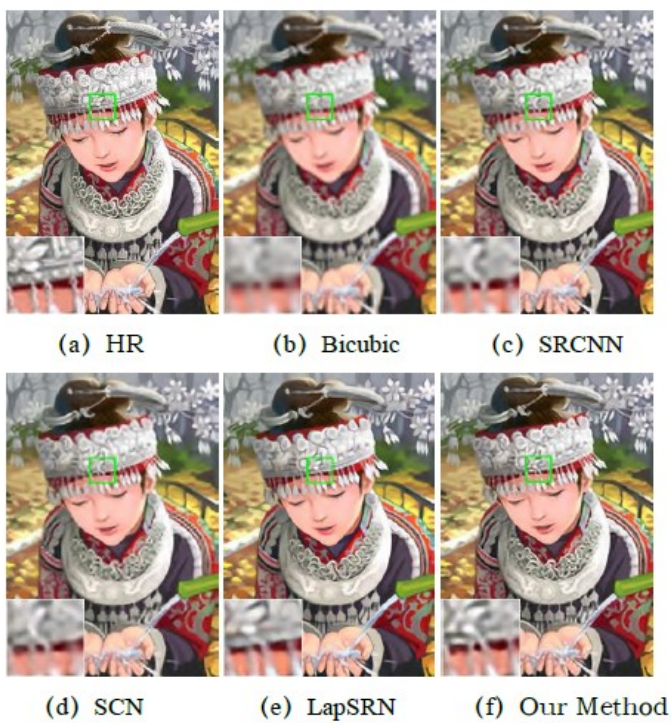

**Figure 5.** Visual quality of different algorithms with the scale factor is 3

**5. Conclusions**

This paper proposes a convolutional neural network based on attention mechanism and multi-scale feature fusion. Through the attention mechanism module, the non-local information and second-order features of the image are effectively fused, the inherent attributes of the image are fully explored, and the rich features are adaptively learned, making our network pay more attention to the information features and improving recognition Learning ability; at the same time, using convolution kernels of different scales to extract the features of the image, and the extracted multi-scale features are fused to preserve the complete information features at multiple scales. The experimental results show that the quality of the image reconstructed by this method is greatly improved in visual and quantitative indicators. The next step is to study a deeper network structure to obtain a more ideal image reconstruction effectiveness.

**References**

1. Zhang, X.; Wu, X., Image interpolation by adaptive 2-D autoregressive modeling and soft-decision estimation. IEEE transactions on image processing 2008, 17, (6), 887-896.
2. Freedman, G.; Fattal, R., Image and video upscaling from local self-examples. ACM Transactions on Graphics (TOG) 2011, 30, (2), 12.
3. Yang J , Wright J , Huang T S , et al. Image Super-Resolution Via Sparse Representation[J]. IEEE Transactions on Image Processing, 2010, 19(11):2861-2873.
4. Dong, C.; Loy, C.C.; He, K.; Tang, X. Learning a deep convolutional network for image super-resolution. In Proceedings of the ECCV 2014, Zurich, Switzerland, 6–12 September 2014.
5. Dong, C.; Loy, C.C.; Tang, X. Accelerating the Super-Resolution Convolutional Neural Network. In Proceedings of the European Conference on Computer Vision 2016, Amsterdam, The Netherlands, 11–14October 2016.
6. Shi, W.; Caballero, J.; Huszár, F.; Totz, J.; Aitken, A.P.; Bishop, R.; Rueckert, D.; Wang, Z. Real-time single image and video super-resolution using an efficient sub-pixel convolutional neural network. In Proceedings of the CVPR 2016, Las Vegas, NV, USA, 27–30 June 2016.
7. Kim, J.; Lee, J.K.; Lee, K.M. Accurate image super-resolution using very deep convolutional networks. In Proceedings of the CVPR 2016, Las Vegas, NV, USA, 27–30 June 2016.
8. Tai, Y.; Yang, J.; Liu, X.; Xu, C. MemNet: A Persistent Memory Network for Image Restoration. In Proceedings of the IEEE International Conference on Computer Vision, Venice, Italy, 22–29 October 2017; pp. 4549–4557.
9. Haris, M.; Shakhnarovich, G.; Ukita, N. In Deep back-projection networks for super-resolution, Proceedings of the IEEE conference on computer vision and pattern recognition, 2018; 2018; pp 1664-1673.
10. Zhang, Y.; Tian, Y.; Kong, Y.; Zhong, B.; Fu, Y. In Residual dense network for image super-resolution, Proceedings of the IEEE Conference on Computer Vision and Pattern Recognition, 2018; 2018; pp 2472-2481.
11. Huang, G.; Liu, Z.; Van Der Maaten, L.; Weinberger, K.Q. Densely Connected Convolutional Networks. In Proceedings of the CVPR 2017 Workshops, Honolulu, HI, USA, 21–26 July 2017; pp. 2261–2269.
12. Tong, T.; Li, G.; Liu, X.; Gao, Q. Image Super-Resolution Using Dense Skip Connections. In Proceedings of the 2017 IEEE International Conference on Computer Vision (ICCV), Venice, Italy, 22–29 October 2017.
13. Timofte, R.; Agustsson, E.; Van Gool, L.; Yang, M.H.; Zhang, L.; Lim, B.; Son, S.; Kim, H.; Nah, S.; Lee, K.M.; et al. Ntire 2017 challenge on single image super-resolution: Methods and results. In Proceedings of the CVPR 2017Workshops, Honolulu, HI, USA, 21–26 July 2017.
14. Bevilacqua, M.; Roumy, A.; Guillemot, C.; Alberi-Morel, M.L. Low-complexity single-image super-resolution based on nonnegative neighbor embedding. In Proceedings of the BMVC 2012, Surrey, UK, 3–7 September 2012.
15. Zeyde, R.; Elad, M.; Protter, M. On single image scale-up using sparse-representations. In Proceedings of the International Conference on Curves and Surfaces, Avignon, France, 24–30 June 2010.
16. Martin, D.; Fowlkes, C.; Tal, D.; Malik, J. A database of human segmented natural images and its application to evaluating segmentation algorithms and measuring ecological statistics. In Proceedings of the ICCV 2001, Vancouver, BC, Canada, 7–14 July 2001.
17. Huang, J.-B.; Singh, A.; Ahuja, N. Single image super resolution from transformed self-exemplars. In Proceedings of the CVPR 2015, Boston, MA, USA, 7–12 June 2015
18. Wang, Z.; Bovik, A.C.; Sheikh, H.R.; Simoncelli, E.P. Image quality assessment: From error visibility to structural similarity. IEEE Trans. Image Process. 2004, 13, 600–612.

19. Kingma, D.; Ba, J. Adam: A method for stochastic optimization. In Proceedings of the ICLR 2015, San Diego, CA, USA, 7–9 May 2015.
20. Lai, W.S.; Huang, J.B.; Ahuja, N.; Yang, M.H. Deep Laplacian Pyramid Networks for Fast and Accurate Super-Resolution. arXiv, 2017; arXiv:1704.03915.